\documentstyle[12pt]{article}
\advance \topmargin by -\headheight
\advance \topmargin by -\headsep

\evensidemargin \oddsidemargin
\def\br{\begin{eqnarray}}
\def\er{\end{eqnarray}}
\def\be{\begin{equation}}
\def\ee{\end{equation}}
\def\lb{\lbrack}
\def\rb{\rbrack}
\def\({\left(}
\def\){\right)}

\relax

\def\a{\alpha}

\def\d{\delta}

\def\g{\gamma}

\def\h{{1\over 2}}
\def\h1p{ {+{1\over 2}}  }

\def\l{\lambda}
\def\lp{l^{\psi}_{a}}

\def\m{\mu}
\def\n{\nu}
\def\o{\over}

\def\p{\phi}
\def\pa{\partial}

\def\tg{\bigtriangleup}
\def\va{\vartheta}

\def\lie{{\cal G}}

\def\rlx{\relax\leavevmode}
\def\inbar{\vrule height1.5ex width.4pt depth0pt}
\def\IZ{\rlx\hbox{\sf Z\kern-.4em Z}}
\def\IR{\rlx\hbox{\rm I\kern-.18em R}}
\def\IC{\rlx\hbox{\,$\inbar\kern-.3em{\rm C}$}}
\def\one{\hbox{{1}\kern-.25em\hbox{l}}}
\begin{document}

\begin{titlepage}

July, 1992 \hfill{IFT-P.26/92}

\hfill{hep-th/9207061}

\vskip .6in

\begin{center}
{\large {\bf Connection between the Affine and Conformal Affine Toda models and
their Hirota's solution}}\footnotemark \footnotetext{Work partially supported
by CNPq}
\end{center}

\normalsize
\vskip .4in

\begin{center}
C. P. Constantinidis\footnotemark
\footnotetext{supported by FAPESP},
L. A. Ferreira, J. F. Gomes,
and A. H. Zimerman
\par \vskip .1in \noindent
Instituto de F\'{\i}sica Te\'{o}rica-UNESP\\
Rua Pamplona 145\\
01405-900 S\~{a}o Paulo, Brazil
\par \vskip .3in

\end{center}

\begin{center}
{\large {\bf ABSTRACT}}\\
\end{center}
\par \vskip .3in \noindent
It is shown that the Affine Toda models (AT) constitute a ``gauge fixed''
version of the Conformal Affine Toda model (CAT).  This result enables one to
map every solution of the AT models into an infinite number of solutions of the
corresponding CAT models, each one associated to a point of the orbit of the
 conformal group.
The Hirota's $\tau$-function are introduced and soliton solutions for the AT
and CAT models associated to $\hat {SL}(r+1)$ and $\hat {SP}(r)$ are
constructed.

\end{titlepage}

\section{Introduction}
Two dimensional integrable non linear models have been studied quite
extensively in the past years. More recently those models possessing conformal
invariance have received special attention. One reason for that is the
connection between integrability and conformal properties, which endow those
theories with a very rich structure.

Among the models most studied are the Toda field theories. These can be
classified in three hierarchies according to the algebraic structure underlying
their associated linear system (zero curvature condition or Lax equation).
First come the Conformal Toda models (CT) associated to the finite simple Lie
algebras with the simplest example being the Liouville model (SL(2)). These are
conformally invariant two dimensional field theories and their solutions have
been constructed using highest weight representations of Lie
algebras~\cite{LS}. The symmetries of these models are described by the so
called W-algebras~\cite{Bilal,Dublin}. The CT theories can be obtained via
Hamiltonian reduction from WZNW models~\cite{Dublin}. The one dimensional
version of these models can also be obtained by Hamiltonian reduction from non
compact symmetric spaces~\cite{Olsha,FO}. Then we have the Affine Toda models
(AT) associated to the loop algebras. These are not conformally invariant but
have been shown to be completly integrable~\cite{Olive1}. Solutions to some
classes of models of this hierarchy have been constructed using different
methods~\cite{leznov,hollo}. However the analysis of Leznov and
Saveliev~\cite{LS} does not work in this case because of lack of highest weight
representations. The most popular member of that hierarchy is the Sinh-Gordon
model associated to the SL(2) loop algebra. Finally there are the recently
proposed Conformal Affine Toda models (CAT) which are related to the Kac-Moody
algebras~\cite{AFGZ,BB,KP}. These are conformally invariant field theories,
and it has been shown that they can be obtained via Hamiltonian reduction from
WZNW models associated to two-loop Kac-Moody algebras~\cite{AFGZ,adao}. The
symmetries of these models are described by some type of W-infinity
algebra~\cite{ACFGZ,W}. The solutions of the CAT models can be constructed
using the Leznov and Saveliev analysis as it has been shown, for the $\hat
{SL}(2)$ case, in ref.~\cite{BB}. Here we give the corresponding solution to
any Kac-Moody algebra.  The solutions of the AT models can then be obtained
from the solutions of the CAT models by setting a particular field to zero.
However, finding the explicit space-time dependence of the solution is quite
hard in pratice due to the fact one is dealing with representations of the
Kac-Moody algebra with non vanishing central term.

The CAT model contains two extra fields with respect to its corresponding AT
counterpart with one of them being a free field.  In this note we show that the
AT models constitute in fact a ``gauge fixed'' version of the CAT models.  The
space of regular solutions of this free field constitute an orbit of the
conformal group.  We then show that the free field can be ``gauged'' away by a
conformal transformation leading the CAT model to the AT model together with
the second extra field, lying in the transformed space time.  As a consequence
of this result any solution of the AT models can be mapped into an infinite
number of solutions of the CAT model, each corresponding to a point of such
orbit.  We also introduce the Hirota's $\tau$ functions for the CAT model and
show how to obtain soliton solutions for the cases of $\hat{SL}(r+1)$ and
$\hat{SP}(r)$.  Our results constitute in fact, a generalization of the
analysis for  $\hat{SL}(r+1)$ given in ref \cite{hollo}.  Explicit examples for
$\hat{SL}(2)$ case are considered.  The generalization for the remaining
Kac-Moody algebras is now under investigation.

\section{The connection between the AT and CAT models}

The equations of motion of the CAT model are given by \cite{AFGZ,BB}:
\br
\pa_{-} \pa_{+} \phi^a &=& \,  q^{a} e^{K_{ab} \phi^b}
- \,q^{0} l^{\psi}_{a}  e^{- K_{\psi b} \phi^b + 2 \mu}
\label{eq:todaone} \\
\pa_{-} \pa_{+} \mu &=& 0 \label{eq:todatwo} \\
\pa_{-} \pa_{+} \nu &=&   {2 \o \psi^2}
\, q^{0} e^{- K_{\psi b} \phi^{b} + 2\mu}
\label{eq:todathree}
\er
where $K_{ab}=2 \a_a.\a_b/{\a_b^2}$ is the Cartan Matrix of a simple Lie
algebra $\lie$, $a,b=1,...,$ rank $\lie$, $\psi$ is the highest root of $\lie$,
$K_{\psi b}=2 \psi .\a_b/{\a_b^2}$, $l^{\psi}_{a}$ are positive integers
appearing in the expansion ${\psi \over
\psi^{2}} = l^{\psi}_{a} {\a_{a} \over \a^{2}_{a}}$, where $\a_a$ are the
simple roots of $\lie$ and $q^a$, $q^0$ are coupling constants. The derivatives
$\pa_{\pm}$ are w.r.t. the light cone coordinates $x^{\pm} = x \pm t$.

These equations can be written in the form of a zero curvature condition
(associated linear system)~\cite{BB,AFGZ}
\be
\pa_{+} A_{-} - \pa_{-} A_{+} + \lb A_{+} , A_{-} \rb = 0
\ee
where
\br
A_{+} = \pa_{+} \Phi + e^{ad \Phi} {\cal E}_{+} \, \, \, , \, \, \,
A_{-} = - \pa_{-} \Phi + e^{-ad \Phi} {\cal E}_{-}
\er
and
\br
\Phi =  {1\over 2} \sum_{a=1}^{rank \lie} \phi^a H_a^0 &+& \mu D + {1\over
2}\nu C  \label{Phi} \\
{\cal E}_{+} = \sum_{a=1}^{rank \lie}  E_{\a_a}^0 +
 E_{-\psi}^1 \, \, \,&,& \, \, \, {\cal E}_{-} = \sum_{a=1}^{rank
\lie} q^a E_{-\a_a}^0 +
q^0 E_{\psi}^{-1}
\er

These equations are invariant under the conformal transformations
\br
x_{+} \rightarrow \tilde{x}_{+} = f(x_{+})
 \, \, \, , \, \, \, \,
x_{-} \rightarrow \tilde{x}_{-} = g(x_{-}) \label{eq:ge}
\er
if the new fields are defined as:
\br
e^{-\tilde{\p}^a(\tilde{x}_+,\tilde{x}_-)} &=& ({df \over dx_+})^{r^{a}}
({dg \over dx_-})^{r^{a}} e^{-\p^a (x_+,x_-)}   \label{eq:fi} \\
e^{-\tilde{\m}(\tilde{x}_+,\tilde{x}_-)} &=& ({df \over dx_+})^{{h \over 2}}
({dg \over dx_-})^{{h \over 2}} e^{-\m (x_+,x_-)}   \label{eq:mi} \\
e^{-\tilde{\n}(\tilde{x}_+,\tilde{x}_-)} &=& ({df \over dx_+})^{B}
({dg \over dx_-})^{B} e^{-\n (x_+,x_-)}
   \label{eq:ni}
\er
where $h$ is the Coxeter number of the algebra $\lie$, $B$ is arbitrary and
$r^a$ is defined as
\be
r^a = \sum ^{rank \lie}_{a=1} K^{-1}_{ab}  \label{eq:ra}
\ee
and it satisfies
\br
\sum^{rank \lie}_{b=1} K_{ab}r^b  = 1 \, \, \, , \, \, \,
\sum^{rank \lie}_{b=1} K_{\psi b}r^b  &=& h - 1      \label{eq:sum2}
\er
Therefore the exponential of the (negative of) fields $\p^a, \m , \n $ are
primary fields of dimensions $ (r^a,r^a), (h/2,h/2), (B,B) $ respectively.

The general solution of the CAT model associated to $\hat {SL}(2)$ was
constructed in ref
\cite{BB} using the method of Leznov and Saveliev \cite{LS}. The generalization
to any other algebra is quite straightforward and the result is
\br
e^{-\phi^a(x_+\, , \, x_{-}) - \lp \nu (x_+\, , \, x_{-})} &=& \langle \l_{(a)}
\mid e^{K_{+}(x_{+})}M_{+}(x_{+}) M^{-1}_{-}(x_{-}) e^{-K_{-}(x_{-})} \mid
\l_{(a)} \rangle
\label{solution1} \\
e^{-\n (x_+\, , \, x_{-})} &=& \langle \l_{(0)} \mid
e^{K_{+}(x_{+})}M_{+}(x_{+}) M^{-1}_{-}(x_{-}) e^{-K_{-}(x_{-})} \mid \l_{(0)}
\rangle
\label{solution2}
\er
and since $\m$ is a free field
\be
\m (x_+\, , \, x_{-}) = \m_{+}(x_{+}) + \m_{-}(x_{-})
\label{solution3}
\ee
where $\m_{\pm}(x_{\pm})$ are arbitrary functions, $K_{\pm}(x_{\pm})$ are
elements of the Cartan subalgebra of the Kac-Moody algebra $\hat{\lie}$
associated to $\lie$, containing the parameters of the solution
\be
K_{\pm}(x_{\pm}) = \sum_{a=1}^{rank \lie} \theta_{\pm}^a(x_{\pm}) H_a^0 \mp 2
\m_{\pm}(x_{\pm}) D + \xi_{\pm}(x_{\pm}) C
\ee
$M_{\pm}$ are exponentiations of real linear combinations of the
positive/negative root step operators of $\hat{\lie}$. The parameters in
$M_{\pm}$ are functions of $x_{+}$/$x_{-}$ only and are determined in terms of
the parameters of the solution through
\br
\pa_{+} M_{+} M_{+}^{-1} &=& - e^{-ad K_{+}} {\cal E}_{+}  \\
M_{-}\pa_{-} M_{-}^{-1} &=& e^{-ad K_{-}} {\cal E}_{-}
\er
The group elements $M_{\pm}$ appear in fact in the Gauss-type decomposition
\be
g_1 = e^{K_{-}}N_{+} M_{-} \, \, \, , \, \, \, g_2 = e^{K_{+}}N_{-}M_{+}
\ee
where $g_1$ and $g_2$ are defined as
\be
e^{-2\Phi} = g_2 g_1^{-1}
\ee
The states $\mid \l_{(a)} \rangle$, $a=1,2,...rank \lie$ and $\mid
\l_{(0)}\rangle$ are highest weight states of representations of the
Kac-Moody algebra ${\hat \lie}$ where the  highest weights are the
fundamental weights $\l_{(a)}$ and $\l_{(0)}$ of ${\hat \lie}$.

Eqs. (\ref{solution1})-(\ref{solution3}) constitute the general
solution for the CAT model equations of motion. We see it depends upon
$2($rank$\lie +2)$ chiral parameters. However, in practice, such result is not
very useful when one wants to know the explicit space-time dependence of a
given solution. The reason is, basically,  that the quantities $M_{\pm}$ are
exponentiations of an infinite number of generators. Unlike the finite
dimensional case \cite{LS} they do not belong to a nilpotent subgroup.  On the
other hand, such form of the solutions are very closely related to dressing
transformations, exchange algebras and are useful in the study of the
symmetries of the model \cite{bernard,Bonora}.

We now show that the Affine Toda model can be understood as a CAT model when
the conformal symmetry is in some sense ``gauge fixed''. The idea consists
basically in conformally tranforming the field $\m$ away for every solution
of it. We start by redefining the fields as
\be
\varphi^a = \p^a - {2r^a \over h} \m  \;\;\;\;\;\;\;  \eta = {2 \over h} \m
\label{eq:eta}
\ee
In terms of them the equations of motion become
\br
\pa_{-} \pa_{+} \varphi^a &=& \,( q^{a}e^{K_{ab} \varphi^b}
- \,l^{\psi}_{a}q^{0}e^{- K_{\psi b} \varphi^b})e^{\eta}
\label{eq:newone} \\
\pa_{-} \pa_{+} \eta &=& 0 \label{eq:newtwo} \\
\pa_{-} \pa_{+} \nu &=&   {2 \o \psi^2}
\, q^{0}e^{- K_{\psi b} \varphi^{b}}e^{\eta}  \label{eq:newthree}
\er
and the quantity $\Phi$ in (\ref{Phi}) becomes
\be
\Phi =  {1\over 2}\left( \sum_{a=1}^{rank \lie} \varphi^a H_a^0 + \eta T_3 +
\nu C \right)
\ee
where $T_3 = 2 {\hat \delta}. H^0 + h D$, with ${\hat \delta}={1\over
2}\sum_{\a > 0}{\a \over {\a^2}}$, is the generator used to perform the
so called homogeneous grading of a Kac-Moody algebra.  Notice that
$e^{\varphi^a}$ are scalars under conformal transformations . If we set the
parameter $B$ to zero in (\ref{eq:ni}), $e^{ \nu} $ is also scalar. On the
other hand $e^{ \eta} $ is a $(1,1)$ primary field. Performing now a
conformal transformation (\ref{eq:fi})-(\ref{eq:ni}) with
\be
f^{\prime}(x_{+}) = e^{\eta_{+}(x_{+})} \, \, \, , \, \, \,
g^{\prime}(x_{-}) = e^{\eta_{-}(x_{-})}
\label{etatransf}
\ee
where $\eta_{\pm}(x_{\pm})$ are solutions of the $\eta$ field, i.e., $\eta
(x_{+},x_{-}) = \eta_{+}(x_{+}) + \eta_{-}(x_{-})$ (see (\ref{solution3})), one
obtains
\br
e^{-\tilde{\varphi}^a(\tilde{x}_+,\tilde{x}_-)} \rightarrow  e^{-\varphi^a
(x_+,x_-)} \, \, , \, \,
e^{-\tilde{\eta}(\tilde{x}_+,\tilde{x}_-)}
\rightarrow   1  \, \, , \, \,
e^{-\tilde{\n}(\tilde{x}_+,\tilde{x}_-)} \rightarrow   e^{-\n (x_+,x_-)}
\label{away}
\er
Therefore the space of regular solutions of the $\eta$ field constitute just
one orbit of the conformal group. Consequently, for every solution of $\eta$,
the equations of motion of the CAT model can be written as
\be
\tilde \pa_{-}  \tilde \pa_{+} \tilde \varphi^a = \,
q^{a}e^{K_{ab} \tilde \varphi^{b}} - \,l^{\psi}_{a}q^{0}e^{-
K_{\psi b} \tilde \varphi^{b}}
\label{eq:decoup1}
\ee
\be
\tilde \pa_{-} \tilde \pa_{+} \tilde \nu =   {2 \o \psi^2}
\, q^{0}e^{- K_{\psi b} \tilde \varphi^{b}}  \label{eq:decoup2}
\ee
where the new space time coordinates are determined in terms of the old ones
through the given solution of $\eta$, i.e. $\tilde x_{+} = \int^{x_{+}}
d{y}_{+} e^{\eta_{+} ({y}_{+})}$, $\tilde x_{-} = \int^{x_{-}} d{y}_{-}
e^{\eta_{-} ({ y}_{-})}$.  From now on we drop the tildes.

Eqs. (\ref{eq:decoup1}) correspond to the Affine Toda equation, which for $\lie
= SL(2)$ become the Sinh-Gordon equation. However
(\ref{eq:decoup1})-(\ref{eq:decoup2}) can be written in a more compact form.
Introduce the fields
\be
\zeta^a = \varphi^a + \lp {\psi^2 \over 2} \n \, \, \, , \, \, \,
\zeta^0 = {\psi^2 \over 2}  \n
\label{defzeta}
\ee
Using the fact that $K_{ab} l^{\psi}_{b} = 2 \a_a . \psi / {\psi^2} \equiv -
K_{a0}$ and $K_{0b} l^{\psi}_{b} \equiv - 2  \psi . \a_b / {\a_b^2}
l^{\psi}_{b} \equiv - K_{00}$ one can write
(\ref{eq:decoup1})-(\ref{eq:decoup2}) as
\be
\pa_{-} \pa_{+} \zeta^i = q^i e^{K_{ij} \zeta^j}
\label{tl}
\ee
where $i,j=0,1,2...rank \lie$, and where we have introduced the extended Cartan
matrix $K_{ij}$ of the Kac-Moody algebra ${\hat \lie}$, $K_{ij} \equiv 2
\a_i . \a_j / {\a_j^2}$ with $\a_i$ being the simple roots of  ${\hat
\lie}$, i.e. $\a_0 = - \psi$ and the remaining $\a_a$, $a=1,2,...rank \lie$, as
before. Eq. (\ref{tl}) is not just the Affine Toda model since it possesses one
extra field. In the literature  one finds the Affine Toda (or Toda Lattice)
model written with one extra field as \cite{Olive1}
\be
\pa_{-} \pa_{+} \rho^i = \sum_{j=0}^{rank \lie}K_{ij} q^j e^{ \rho^j}
\label{tlo}
\ee
The usual Affine Toda is then obtained  by noticing that the extended Cartan
matrix is singular and therefore has a null vector, $ n^i K_{ij} = 0$ where
$\psi = n_a^{\psi} \a_a$ and $n_0^{\psi} =1$. So, the field $\rho \equiv
n_i^{\psi} \rho^i$ satisfies $\pa_{-} \pa_{+} \rho = 0$. Restricting the model
to the vacuum $\rho = 0$ one eliminates one field and obtains the usual Affine
Toda as given by (\ref{eq:decoup1}) \cite{Olive1}. However going from
(\ref{tl}) to (\ref{tlo}) would involve a singular transformation between the
fields $\zeta^i$ and $\rho^i$ and therefore they are not really the same model.

The conformal transformation (\ref{etatransf}) is closely related to the
transformation performed in ref. \cite{adao} on the currents of the two-loop
Kac-Moody algebra \cite{AFGZ} in order to decouple the non-zero modes of a
central current . Under (\ref{etatransf}), the two-loop Kac-Moody currents
$J_R(x_{+})\equiv k {\hat g}^{-1} \pa_{+} {\hat g}$ and $J_L(x_{-})\equiv -k
\pa_{-} {\hat g} {\hat g}^{-1}$ , where ${\hat g}$ is an exponentiation of
generators of an ordinary Kac-Moody algebra \cite{AFGZ,adao}, transform as
\be
J_R(x_{+}) \rightarrow e^{\int^{x_{+}}d{ y}_{+} C_R({ y}_{+})}
J_R(x_{+})
\, \, \, \, , \, \, \, \,
J_L(x_{-}) \rightarrow e^{\int^{x_{-}}d{y}_{-} C_R({y}_{-})}
J_L(x_{-}) \label{J}
\ee
where $C_{R/L}$ are the components of the currents in the direction of the
generator $D$ and can be written as $C_{R/L}(x_{\pm}) = \pa_{\pm}
\eta_{\pm}(x_{\pm})$ with $\eta_{\pm}$ as in (\ref{etatransf}). These are the
transformations performed in \cite{adao} except that there the zero modes of
$C_{R/L}$ were excluded due to the periodic boundary condition imposed on
$J_{R/L}$.

\section{Hirota's method}
We now describe how to use the Hirota's method \cite{Hirota} to construct
solutions for the Affine and Conformal Affine Toda models. We introduce the
$\tau$-functions as
\be
\zeta^j = l_j^{\psi}( - \ln \tau_j +  \sigma ) + \va_j
\label{tau}
\ee
where $l_0^{\psi} = 1$ and $\lp$, $a=1,2,... rank \lie$ as before and
\be
\va_0 = 0 \, \, \, \, , \, \, \, \,
\va_a = \sum_{b=1}^{rank \lie} (RK)^{-1}_{ab} \ln \left( {q^0 l_b^{\psi} \over
q^b} \right)
\ee
where $R$ is the matrix with entries $R_{ab}=\delta_{ab} + n_b^{\psi}$,
with $n_b^{\psi}$ being the integers in the expansion $\psi =
n_a^{\psi} \a_a$, $ l^{\psi}_{a} =  {\a^{2}_{a}\o {\psi ^2}}n_a^{\psi}$  and so
\be
(R^{-1})_{ab} = \delta_{ab} - {n_b^{\psi} \over h}
\label{niceinv}
\ee
where $h=\sum_{a=1}^{rank \lie} n_a^{\psi} + 1$ is the Coxeter number of
$\lie$. The $\va_a$ are associated to the vacuum of the fields $\varphi_a$ and
were introduced to make Hirota's equations independent of the coupling
constants $q^j$. We have the freedom to introduce the field $\sigma$ because it
will drop from the exponential interaction term, since $l_i^{\psi}$ is a null
vector of the extended Cartan matrix $K_{ij}l_j^{\psi} = 0$. In addition,  it
is related to the vacuum of the field $\n$, and it will help us in decoupling
Hirota's equations into a suitable form.

To simplify the notation we introduce the operator
\be
\bigtriangleup (F) \equiv \pa_{+} \pa_{-} \ln F = {\pa_{+} \pa_{-} F \over F} -
{\pa_{+} F \pa_{-} F \over F^2}
\label{triangle}
\ee
which obviously satisfies
\be
\tg (FG) = \tg (F) + \tg (G) \, \, \, , \, \, \, \tg (F^{\gamma}) = \gamma \tg
(F)
\label{proptg}
\ee
The equations one obtains by substituting (\ref{tau}) into (\ref{tl}) can be
decoupled as
\br
\tg (\tau_j) &=& - \beta \left(  \prod_{k=0}^{rank \lie}
\tau_k^{-K_{jk}l_k^{\psi}} - 1 \right) \label{tau1}\\
\pa_{+} \pa_{-} \sigma &=& \beta \label{tau2}
\er
where
\be
\beta = {q^j \over l_j^{\psi}} e^{K_{jk}\va_k} \, \, \, \, \, \, \, \,
\mbox{for any $j=0,1,...,rank \lie $}
\ee
which, using (\ref{niceinv}), one can easily show that it is a constant
independent of the index $j$. The solution for $\sigma$ is therefore
\be
\sigma (x_{+}, x_{-}) = \beta x_{+} x_{-} + F(x_{+}) + G(x_{-})
\ee
with $F$ and $G$ arbitrary functions. From (\ref{defzeta}) we get the relation
between $\tau$-functions and the original CAT model fields
\be
\varphi^a = -l_a^{\psi} \ln {\tau_a \over \tau_0} + \va_a \;\;\;\;\;\;\;\;
\n = {2 \over {\psi}^2} \left( \sigma - \ln \tau_0 \right) \label{phinu}
\ee
In ref \cite{hollo} the Hirota's solution for the Affine Toda models associated
to $\hat{SL}(r+1)$ was constructed. In order to get the Hirota's equation the
number of $\tau$-functions introduced exceeded the number of fields by one. It
is now clear that the extra $\tau$- function, namely $\tau_0$, corresponds to
the  $\n$ field and that the Hirota's equation is intrinsically related to the
structure of the CAT model.

Notice that the Hirota's equation (\ref{tau1}) is invariant under scale
transformations on the $\tau$-functions, $\tau_j \rightarrow \lambda \tau_j$,
since $l_j^{\psi}$ is a null vector of the extended Cartan matrix. This
guarantees that when both sides of (\ref{tau1}) are multiplied by
$\tau_j^{K_{jj}l_j^{\psi}}$, all terms will be of the same order
in $\tau$. Obviously, the equations will be bilinear in $\tau$
only for those algebras where $l_j^{\psi}=1$ for all $j$'s. That
happens for the algebras ${\hat SL}(r+1)$ and ${\hat Sp}(r)$.
We next apply the Hirota method to the CAT model described by
(\ref{eq:decoup1}) and (\ref{eq:decoup2}) for these two
algebras. The application of such method to the remaining
algebras is now under investigation and it will be published
elsewhere.

\subsection{ $\hat{SL}(r+1)$}

For $\hat{SL}(r+1)$ the N soliton solution from Hirota's method
was obtained in \cite{hollo}.  The $\tau$ functions for the
$\hat{SL}(2)$ CAT and AT models were also considered in
ref.\cite{bernard,faddeev}.   Here we give a short account of those
results for completeness.  The integers $l^{\psi}_{a},
a=1,2,...r $ are all equal to unity in this case.  The extended
Cartan matrix is obtained from the usual one by adding an extra
row and columm corresponding to an extra point in the Dynkin
diagram.  It is given by
\be
K= \left( \begin{array}{rrrrrrr}
2 & -1 & 0 & 0 & \ldots & 0 & -1 \\
-1 & 2 & -1 & 0 & \ldots & 0 & 0 \\
0 & -1 & 2 & -1 & \ldots & 0 & 0 \\
\vdots & \vdots & \vdots & \vdots & \vdots & \vdots & \vdots \\
0 & 0 & \ldots & 0 & -1 & 2 & -1 \\
-1 & 0 & 0 & \cdots & 0 & -1 & 2 \end{array} \right)
\ee
The equations for the tau functions corresponding to (\ref{tau1}) becomes
\be
\tau_j^2 \tg (\tau_j) = \beta \left( \tau_j^2 -  \tau_{j+1} \tau_{j-1}\right)
\ee
$j=0,1,...r$, where $\tau_{j+r+1} = \tau_{j}$ is understood from
the periodicity of the extended Dynkin diagram.
The above system admits N soliton solutions of the form
\be
\tau_j = 1 + \epsilon \tau^{(1)}_j + ... + \epsilon^{N} \tau^{(N)}_j
\label{Nsoliton}
\ee
We now seek for one soliton solution, i.e. $ \tau_j = 1 + \epsilon
\tau^{(1)}_j $, where
\be
\tau^{(1)}_j = \exp(\g (x-vt-\xi ))\delta_j \label{tau1j}
\ee
The truncation of (\ref{Nsoliton}) in order $\epsilon $ is consistent if we
impose the following recursion relations
\be
\delta_j^{2} = \delta_{j-1} \delta_{j+1}
\ee
An obvious solution satisfying the periodicity condition, $\tau^{(1)}_{r+1} =
\tau^{(1)}_0 $ is given in terms of the $(r+1)$-th root of unit, i.e.
\be
\delta_j = \omega^j
\ee
where $\omega = \exp({{2\pi ik}\over {r+1}}), k=0,1,2,...r$.  The wave
parameters $\g $ and $v$ are related by  ($4 \pa_- \pa_+ = \pa_x^2 -\pa_t^2$)
\be
\g^2 (1 - v^2) = 16\beta \sin^2 ({k\pi \over {r+1}})
\ee

\subsection{$\hat{SP}(r)$}
Again in this case the integers $l^{\psi}_j = 1, j=0,1,...r$.  The extended
Cartan matrix is given by
\be
K= \left( \begin{array}{rrrrrrr}
2 & -2 & 0 & 0 & \ldots & 0 & 0 \\
-1 & 2 & -1 & 0 & \ldots & 0 & 0 \\
0 & -1 & 2 & -1 & \ldots & 0 & 0 \\
\vdots & \vdots & \vdots & \vdots & \vdots & \vdots & \vdots \\
0 & 0 & \ldots & 0 & -1 & 2 & -1 \\
0 & 0 & 0 & \cdots & 0 & -2 & 2 \end{array} \right)
\ee
The system of equations for the tau functions corresponding to (\ref{tau1}) can
therefore be obtained by reading off the elements of the extended Cartan
matrix, i.e.
\br
\tau_0^{2} \tg (\tau_0) &=& \beta \left( \tau_0^{2} -
\tau_1^{2}\right) \nonumber\\
\tau_a^{2}\tg (\tau_{a}) &=& \beta \left( \tau_a^{2} -
\tau_{a+1} \tau_{a-1}\right) \, \, \, \, \, \, \, \mbox{for
$a=1,2,...,r-1$} \nonumber\\
\tau_r^{2}\tg (\tau_r) &=& \beta \left(\tau_r^{2} -  \tau_{r-1}^{2}\right)
\er
The one soliton solutions is again obtained setting  $ \tau_j = 1 + \epsilon
\tau^{(1)}_j $, where
\be
\tau^{(1)}_j = \exp(\g (x-vt-\xi ))\delta_j \label{exp}
\ee
This case leads to two solutions.  One is obtained setting $\d_0 = \d_1 = ... =
\d_r $ yielding $\g \neq 0$ and $v^2 =1 $, i.e.,
\be
\tau^{(1)}_j =  \exp{(\g (x \pm t -\xi ))}\d_0
\ee
implying from (\ref{phinu})
\br
\varphi^a = \va_a \, \, \, \, , \, \, \, \,
\nu = {2 \o \psi^2} \left( - \ln \left( 1 + \exp{(\g (x \pm t -\xi ))}\d_0
\right) + \beta x_{+} x_{-} + F(x_{+}) + G(x_{-}) \right)
\er
The second solution is obtained when $\d_0 = \d_2 = ... = \d_{2n}$ and
$-\d_0 = \d_1 = \d_3 = ... =\d_{2n+1}$ yielding
\be
\tau^{(1)}_j = (-1)^j \exp{(\g (x -vt -\xi ))}\d_0
\ee
where  the wave parameters are related as
\be
\g^2 (1 - v^2) = 16\beta \d_0
\ee
Again from (\ref{phinu}) we get
\br
\varphi^{a} &=& \va_{a} \, \, \, \, \, \mbox{for $a$ even} \nonumber\\
\varphi^{a} &=& - \ln \left( {{1 - \exp{(\g (x -vt -\xi ))}\d_0 } \o
{1 + \exp{(\g (x -vt -\xi ))}\d_0 }} \right) + \va_{a}
\, \, \, \, \mbox{for $a$ odd} \nonumber\\
\nu &=& {2 \o \psi^2} \left( - \ln \left( 1 + \exp{(\g (x - v t -\xi ))}\d_0
\right) + \beta x_{+} x_{-} + F(x_{+}) + G(x_{-}) \right)
\er
where $a=1,2,...r$. Notice that $\varphi^{a}$ for $a$ odd constitute copies of
the Sinh-Gordon soliton.

\subsection{The $\hat{SL}(2)$ Case}

We now illustrate the procedure described in the previous sections where
the solutions of the CAT model were shown to be related to those of the AT
model with space time variables parametrized in terms of a free field $\eta $
as follows
\br
{\tilde x} & = &{1 \o {2}}({\tilde x}_{+} + {\tilde x}_{-}) = {1 \o
{2}}(\int^{x_{+}} dy e^{\eta_{+} (y)} + \int^{x_{-}}d{y} e^{\eta_{-}
({y})}) \nonumber \\
{\tilde t} & = &{1 \o {2}}({\tilde x}_{+} - {\tilde x}_{-}) = {1 \o
{2}}(\int^{x_{+}} d{y} e^{\eta_{+} ({y})} - \int^{x_{-}}d{y} e^{\eta_{-}
({y})})
\label{newst}
\er

Let us consider the CAT model described by equations
(\ref{eq:newone})-(\ref{eq:newthree}) for ${\hat SL}(2)$.  The solution given
in terms of the $\tau $-functions in (\ref{phinu}) reads
\br
\varphi (x,t)&=& -\ln \left({{1 - i\exp (\g ({\tilde x}-v{\tilde
t}))}\o {1 + i\exp (\g ({\tilde x}-v{\tilde t}))}}\right) +
{1\o{4}}\ln{q^0 \o {q^1}}\nonumber \\
&=& -2i\arctan (\exp (\g ({\tilde x} - v{\tilde t}))) +  {1\o {4}}\ln{q^0 \o
{q^1}}\nonumber \\
\nu (x,t)&=& \sigma - \ln\bigl[1 + i\exp (\g ({\tilde x}-v{\tilde
t})\bigr] \nonumber \\
\eta (x,t)& =& \eta_+(x+t) + \eta_-(x-t)
\er
where we have made the choice $\xi = {\pi i \o 2\g}$ in (\ref{tau1j}).
Equations (\ref{newst}) stress the fact that there are  infinite many space
time new variables in correspondence with the choice of $\eta_{\pm}$.  In
particular for $\eta_+ =
\eta_- = 0$ the solution for the CAT model field $\varphi$ is the same as that
for the Sinh-Gordon field .

Consider, for instance, the usual static (in ${\tilde t}$) soliton solution for
the Sinh-Gordon model, i.e. $v=0$ \cite{Rajaraman}. The corresponding solution
for the ${\hat {SL}}(2)$ CAT model for given $\eta_{\pm}$ will not in general
be static (in $t$). In fact, the solution can be completly different and in
some cases the solitonic character is lost.  This means its topological charge
may not be preserved by the conformal transformation. We have checked that for
$\eta_{\pm} = x_{\pm}$ the static Sinh-Gordon soliton is mapped into a
solitonic solution of the ${\hat {SL}}(2)$ CAT model. It travels preserving its
asymptotic behaviour, and so having a topological charge. One can describe its
trajectory by giving, for instance, the position of the point $\varphi = {3 \pi
\over 4}$.
\be
x = - \ln (\cosh t) + x_0
\ee
So, it travels with velocity ${dx \o dt}= - \tanh t$ coming from $x=-\infty$
(for $t=-\infty$) bouncing at the point $x_0=\ln({1 \o {\g}}\ln( \tan ({3\pi \o
{4}} -{1\o{4}}\ln{q^0 \o {q^1}})) $ and then returning to $x=-\infty$ (for
$t=\infty$).

Such procedure can be used to study the perturbation caused by the $\eta$ field
on the solitons of the Affine Toda models. It would be interesting to develop a
systematic way of doing that.

\end{document}